\documentclass[preprint]{aastex61}

\newcommand{\rsi}{Rev. Sci. Instr.}

\usepackage{tablefootnote}
\usepackage{threeparttablex}

\begin{document}

\title{High-resolution charge exchange spectra with L-shell nickel show striking differences from models}

\correspondingauthor{G. L. Betancourt-Martinez}
\email{Gabriele.Betancourt@irap.omp.eu}

\author{G. L. Betancourt-Martinez}
\affil{IRAP CNRS \\
9 Av. du Colonel Roche, BP 44346 \\
31028 Toulouse Cedex 4, France}

\author{P. Beiersdorfer}
\affiliation{Lawrence Livermore National Laboratory \\
Livermore, CA, 94550 USA}

\author{G. V. Brown}
\affiliation{Lawrence Livermore National Laboratory \\
Livermore, CA, 94550 USA}

\author{R. S. Cumbee}
\affiliation{NASA Goddard Space Flight Center \\
Greenbelt, MD 20771 USA}
\affiliation{Universities Space Research Association \\
Columbia, MD 21046 USA}

\author{N. Hell}
\affiliation{Lawrence Livermore National Laboratory \\
Livermore, CA, 94550 USA}

\author{R. L. Kelley}
\affiliation{NASA Goddard Space Flight Center \\
Greenbelt, MD 20771 USA}

\author{C. A. Kilbourne}
\affiliation{NASA Goddard Space Flight Center \\
Greenbelt, MD 20771 USA}

\author{M. A. Leutenegger}
\affiliation{NASA Goddard Space Flight Center \\
Greenbelt, MD 20771 USA}
\affiliation{CRESST and University of Maryland Baltimore County \\
Baltimore, MD 21250 USA}

\author{T. E. Lockard}
\affiliation{Lawrence Livermore National Laboratory \\
Livermore, CA, 94550 USA}

\author{F. S. Porter}
\affiliation{NASA Goddard Space Flight Center \\
Greenbelt, MD 20771 USA}

\begin{abstract}

We present the first high-resolution laboratory spectra of X-ray emission following L-shell charge exchange between nickel ions and neutral H\textsubscript{2} and He. We employ the commonly-used charge exchange models found in \textsc{xspec} and \textsc{spex}, \textsc{acx} and \textsc{spex-cx}, to simulate our experimental results. We show that significant differences between data and models exist in both line energies and strengths. In particular, we find that configuration mixing may play an important role in generating lines from core-excited states, and may be improperly treated in models. Our results indicate that if applied to astrophysical data, these models may lead to incorrect assumptions of the physical and chemical parameters of the region of interest.

\end{abstract}

\keywords{atomic processes --- line: formation --- methods: laboratory: atomic}


\section{Introduction}\label{sec:pap_intro}

Charge exchange (CX) is the radiationless transfer of one or more electrons from a neutral atom or molecule to an excited state of a highly charged ion and the subsequent radiative de-excitation, and results in characteristic X-rays. CX has been found to be an important mechanism for spectral line formation in the solar system: solar wind CX is observed around comets \citep{1996Sci...274..205L,1997GeoRL..24..105C}, planetary exospheres, including our own \citep{2012AN....333..324D}, and it is a significant contributor to the soft X-ray background \citep{2014Natur.512..171G}. CX has also been postulated to occur astrophysically, in supernova remnants \citep{2014ApJ...787L..31C, 2015MNRAS.449.1340R}, clusters of galaxies \citep{2015MNRAS.453.2480W, 2017ApJ...837L..15A}, stellar winds \citep{2007A&A...463.1111P}, and galactic winds \citep{2007PASJ...59S.269T}. 

Most observational and modeling efforts for CX have concentrated on K-shell ions. However, L-shell ions make up a non-negligible abundance fraction in a wide variety of X-ray sources, such as in Jupiter's polar regions \citep{2002Natur.415.1000G}, and the solar wind \citep{2000ApJ...544..558S}. In particular, L-shell Ni ions, though less cosmically abundant than Fe, have been identified in spectra of stellar coronae \citep{2001ApJ...548..966B, 2009A&ARv..17..309G, 2015A&A...577A..93P} as well as in high-resolution spectra of the Sun \citep{1982ApJ...256..774P}. As high-resolution spectral measurements become more routine, it will become increasingly important to understand the behavior of Ni L-shell lines compared to neighboring L-shell Fe lines in order to properly interpret spectral line diagnostics.  

Because of the significance of CX in astrophysics, including its ability to quickly lower the charge state of a plasma and/or significantly alter its assumed chemical abundance, CX models are becoming more readily available. For example, \textsc{acx} \citep{2014ApJ...787...77S} and \textsc{spex-cx} \citep{2016A&A...588A..52G}, available in the \textsc{xspec} and \textsc{spex} spectral modeling packages, respectively, have become popular tools to test for the presence of CX or explain anomalous X-ray emission in astrophysical spectra. We rely on the accuracy of these and other models and the atomic databases at their cores to perform and understand our scientific analyses. However, recent results have shown that certain comparisons across models may yield dramatically differing results \citep{2017arXiv171205407H}. For the case of CX, experimental data are often in conflict with models, even for K-shell ions (e.g., \cite{2000PhRvL..85.5090B, 2003Sci...300.1558B, 2005ApJ...634..687W, 1742-6596-58-1-032, 2010PhRvL.105f3201L}). The situation is worse for CX onto L-shell ions due to their more complex atomic structure: few experimental spectra of L-shell CX exist, especially at high resolution (e.g., \cite{1979PhRvA..19..504C, 1985JPhB...18.4763D, 1992JPhB...25.3009S, 1995PhRvA..51.3685F, 2000JPhB...33.5275L, 2001PhRvL..86..616L,2000AIPC..500..626B, 2002PhRvA..65d2509T, 2003NIMPB.205..605T,2009ApJ...702..171F}), and results from comparisons of these experiments to theoretical models, ranging from classical to quantum mechanical, are mixed, but tend towards agreeing at high collision energies and showing significant discrepancies at the low collision energies relevant to astrophysics \citep{1992JPhB...25.3009S, 2000JPhB...33.5275L, 2001PhRvL..86..616L}. 

By performing laboratory experiments of CX between various ion and neutral species, we can learn more about the detailed atomic physics of CX and assess the accuracy and limitations of our models, determining at the atomic level where any uncertainties may exist. In this paper, we present recent measurements of CX between Ni$^{19+}$ and He and H\textsubscript{2} with an Electron Beam Ion Trap (EBIT) and an X-ray microcalorimeter, and compare our experimental results to spectra produced by \textsc{spex-cx} and \textsc{acx}. We show that there are disconcerting differences between the experimental and model spectra across the L-shell Ni energy band that may stem from inconsistencies across or inaccuracies in the atomic databases.  

\section{Experimental Method and Line Identification}

For our experiments, we used the EBIT-I electron beam ion trap at the Lawrence Livermore National Laboratory (LLNL) \citep{2003NIMPB.205..173B} and measured the spectra with the EBIT Calorimeter Spectrometer (ECS) \citep{Porter08}. The general experimental method is described in \cite{2014PhRvA..90e2723B}. In brief, we create and trap our ion of interest with a tunable electron beam (a stage called direct excitation, or DE), then turn off the beam and magnetically confine the ions while we inject our neutral species, allowing CX to occur. The injection of the ion and neutral species is continuous. The ions are then dumped and the cycle is repeated, often for several hours to days to collect sufficient counts. For the experiments presented here, the length of the DE and CX phases in our analyses were approximately 0.2 and 0.3 seconds, respectively, for a total cycle time of $\sim$0.5 seconds. Nickel was supplied by sublimation of nickelocene (C\textsubscript{10}H\textsubscript{10}Ni) which flowed directly into the EBIT trap region. We tuned the electron beam energy to breed mostly F-like Ni (Ni$^{19+}$), which leads to Ne-like (Ni$^{18+}$) following single electron capture (SEC) in CX. Ne-like Ni was also present during charge breeding. This was necessary in order to avoid creating O-like, thus F-like following CX, which has several spectral lines within the Ne-like band. We injected neutral He and H\textsubscript{2} directly into the trap via a ballistic  gas injector. Typical thermal energies of trapped ions in the EBIT are $\sim$10 eV amu$^{-1}$ ($\sim$50 km s$^{-1}$) \cite{1995RScI...66..303B, 1996PhRvL..77.5353B}. This is the approximate collision energy/velocity at which CX occurs in our experiments.

The ECS is a silicon-thermistor X-ray microcalorimeter which is described in detail in \cite{Porter08}. It has a 30-pixel array of silicon-doped thermistors which are divided into a mid- and a high-energy array. The experiments discussed here made use of 14 pixels in the mid-band array. These have an energy resolution of $\sim$4.5 eV at 6 keV and an absorber quantum efficiency across the Ni L-shell energy band of nearly unity \citep{Porter08}. There are four aluminized polyimide infrared/optical blocking filters in the optical path of the ECS, as described in \cite{2014PhRvA..90e2723B}. In addition, during the experiments we checked for the presence of background contaminants that might have frozen onto one or more of the filters, such as nitrogen or water ice, which reduces the X-ray transmission. The total transmission for the experiments presented here varies smoothly in our band of interest from 0.71 at 880 eV to 0.92 at 1499 eV. The energy scale was calibrated for each pixel using X-ray emission from H- and He-like ions of O, Ne, S, and Ar, and is accurate to within 0.5 eV.

The data are time-tagged and phase-folded on the EBIT-I cycle time. For the analysis presented here, we used CX data with phase times $\geq$2 ms after the electron beam was turned off. This allows the metastable $1s^{2}_{1/2}2s^{2}_{1/2}2p_{1/2}2p^{4}_{3/2}(J=1/2)$ state of Ni\textsuperscript{19+}, which has a lifetime of $\sim$20 $\mu$s, to relax to $1s^{2}_{1/2}2s^{2}_{1/2}2p_{1/2}^{2}2p^{3}_{3/2}(J=3/2)$. We measured the spectrum and count rate that resulted from CX with background gases in the trap by ceasing injection of our desired neutrals, and we subtracted the background spectrum. We performed these background measurements periodically during the experimental campaign period, at the beginning of the day, after the trap was pumped out overnight. 

To identify the significant spectral lines present in our CX spectra, we used the energy scale and line identifications from the DE spectrum. We then compared the measured line centroids to calculations with the Flexible Atomic Code (\textsc{fac}) version 1.1.3 \citep{2008CaJPh..86..675G} and measurements of L-shell Ni lines with a high-resolution grating spectrometer from \cite{2007ApJ...657.1172G}. In our \textsc{fac} calculation, we corrected the ground state ionization energies of Ne-like Ni according to Scofield (private communication), and the $3 \rightarrow 2$ transition energies to match those in \cite{2007ApJ...657.1172G}. The error on our calculated transition energies is dominated by ground state ionization energies, which we estimate to be $\sim$1 eV. These line identifications are presented in Table \ref{table:lineID}. In some cases, one line may be a combination of several transitions presented. Some F-like Ni lines are also included in our identifications; these would be present following SEC in CX with an O-like ion. They should only be a minor contribution to the spectrum due to the fact that the O-like $2p_{1/2}2p_{3/2}^{2}3d_{5/2}(J=3) \rightarrow 2p_{3/2}^{2}(J=2)$ transition at $\sim$1096 eV, which should be the strongest one in the O-like series in DE \citep{2007ApJ...657.1172G}, is not significant in our spectra. 

The spectra shown in the following sections are background-subtracted but are not corrected for filter attenuation; the models are adjusted for this attenuation to match the experimental spectra. 

\section{Key Spectral Features}\label{sec:spectra}

\begin{figure}[ht]
 \begin{center}
 \includegraphics[scale=0.65, angle=90]{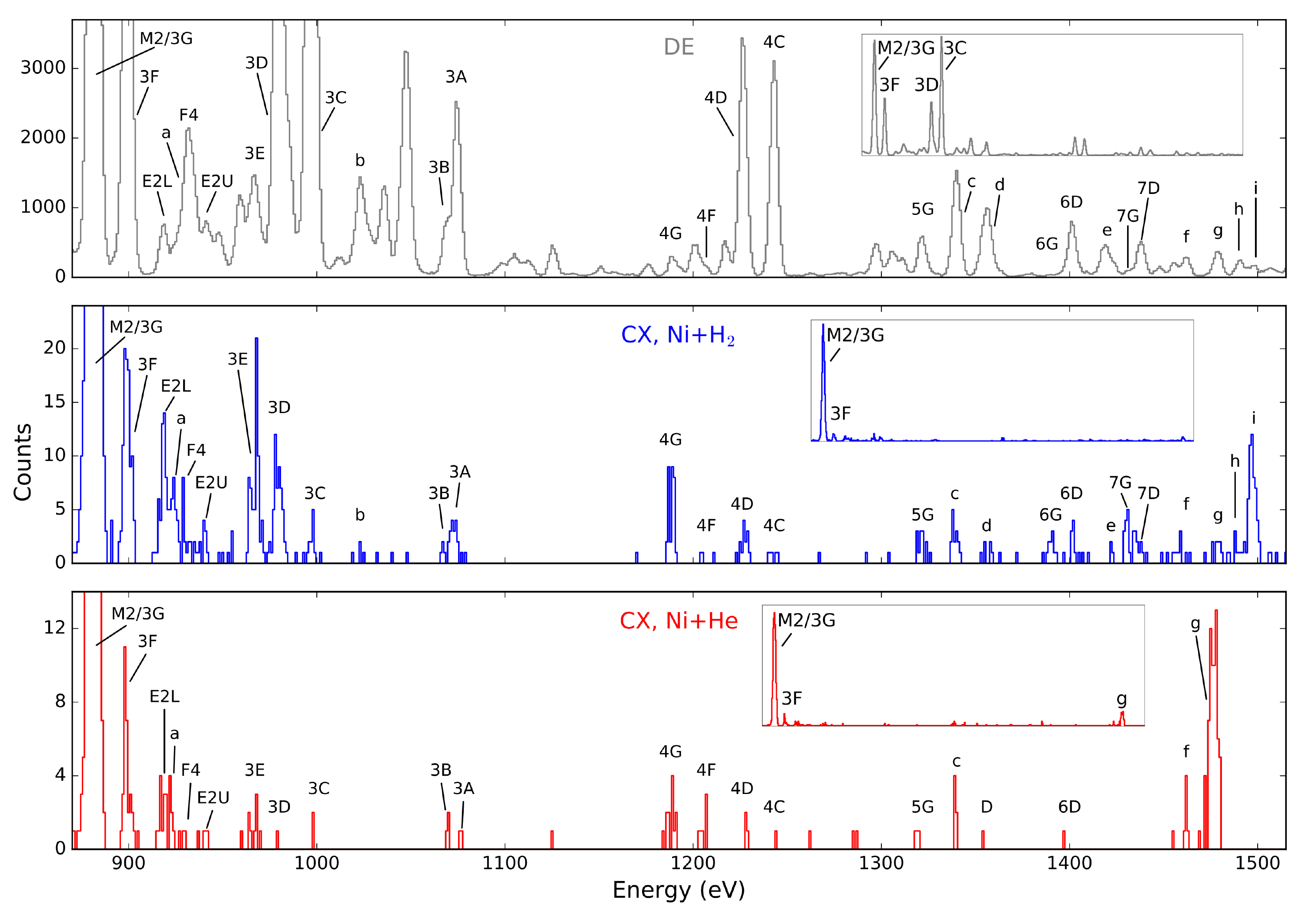}
 \end{center}
 \renewcommand{\baselinestretch}{1}
\small\normalsize
\begin{quote}
\caption{Magnified spectra of Ni$^{19+}$ and Ni$^{18+}$ created in DE (top) and CX between Ni$^{19+}$+H\textsubscript{2} (middle) and Ni$^{19+}$+He (bottom). The full L-shell spectra are shown in the insets. Significant CX lines are identified by labels that correspond to entries in Table \ref{table:lineID}; lines that may result from blends are labeled alphabetically. The M2/3G lines dominate the CX spectra, while the 3F, 3D, and 3C lines are suppressed compared to DE. We also observe a difference in $n$\textsubscript{max} $\rightarrow 2$ transitions from 1450 eV--1500 eV between the Ni$^{19+}$+H$_2$ and the Ni$^{19+}$+He spectra, which likely stems from the differing ionization potentials of the two neutrals. In this and all following figures, the data have been background-subtracted but have not been adjusted for filter transmission. The total number of counts in the M2/3G lines in the DE, Ni$^{19+}$+H\textsubscript{2}, and Ni$^{19+}$+He spectra are $\sim$133000, $\sim$1750, and $\sim$600, respectively.}
\label{fig:allspec} 
\end{quote}
\end{figure}

Figure \ref{fig:allspec} presents our measured CX spectra with both neutral partners, as well as a DE spectrum following collisional excitation. The strongest line in both Ni$^{19+}$+H$_2$ and Ni$^{19+}$+He CX spectra is a blend of the M2 ($2p_{3/2}^33s_{1/2}(J=2) \rightarrow 2p_{3/2}^4(J=0)$) and 3G ($2p_{3/2}^33s_{1/2}(J=1) \rightarrow 2p_{3/2}^4(J=0)$) lines. This is a stark difference from the DE spectrum, which at the $\sim$4.5 eV resolution of the ECS, has four prominent lines from $n=3 \rightarrow 2$ transitions: M2/3G, 3F ($2p_{1/2}^13s_{1/2}^1(J=1) \rightarrow 2p_{3/2}^4(J=0)$), 3D ($2p_{3/2}^33d_{5/2}^1(J=1) \rightarrow 2p_{3/2}^4(J=0)$), and 3C ($2p_{1/2}^13d_{3/2}^3(J=1) \rightarrow 2p_{3/2}^4(J=0)$). The relative enhancement of the M2/3G lines and suppression of the 3F, 3D, and 3C lines may be seen as strongly diagnostic of the presence of CX. 

\begin{ThreePartTable}
\begin{TableNotes}
\footnotesize
\item [a] Line label as shown in the figures in this work, following the notation of \cite{parkinson, loulergue_nussbaumer}, or lowercase alphabetic label if the observed line may result from a blend of transitions or is unidentified. An asterisk indicates a line that results from core excitation.
\item [b] Line identification of a possible component of a blended line. Labels follow the notation of \cite{parkinson, loulergue_nussbaumer}, and an asterisk indicates a line that results from core excitation. A dagger indicates an unidentified line.
\item [c] The energy of the peak of the line measured in our DE spectrum.
\item [d] Line energy calculated in \textsc{fac}.
\item [e] Line energy in \cite{2007ApJ...657.1172G}.
\item [f] Line energy (converted from observed wavelength) from the NIST atomic spectra lines database. 
\item [g] Upper state for the transition presented. The lower state for all Ni$^{18+}$ ions is $2p_{3/2}^{4}(J=0)$, and the lower state for the Ni$^{19+}$ transitions here is $2p_{3/2}^3(J=3/2)$.
\end{TableNotes}
\begin{longtable}{ccccccc}
\caption{Line identifications for the strongest lines observed in our measured CX spectra}
\label{table:lineID} \\
\hline \hline
\textbf{Label}\tnote{a} & \textbf{Secondary Label}\tnote{b} &\textbf{DE (eV)}\tnote{c} & \textbf{FAC (eV)}\tnote{d} & \textbf{Gu (eV)}\tnote{e} & \textbf{NIST (eV)}\tnote{f} & \textbf{Upper State}\tnote{g} \\
\hline \hline
\endfirsthead
M2	&		&	882	&	880.753	&	880.827	&	880.827	&	$2p_{3/2}^{3}3s_{1/2}(J=2)$\\
3G	&		&		&	882.888	&	882.96	&	882.960 	&	$2p_{3/2}^{3}3s_{1/2}(J=1)$\\
3F*	&		&	900	&	899.807	&	899.877	&	899.877 	&	$2p_{1/2}3s_{1/2}(J=1)$\\
E2L	&		&	919	&	920.037	&	918.542	&	 		&	$2p_{3/2}^{3}3p_{1/2}(J=2)$\\
a	&		&	925	&	926.041	&			&	 		&	$2p_{3/2}^{3}3p_{3/2}^{3}(J=2)$\\
F4	&		&	932	&			&	931.586	&	 		&	$2p_{3/2}^{2}3s_{1/2}(J=5/2)$\\
E2U*&		&	941	&	942.002	&	940.346	&	 		&	$2p_{1/2}3p_{3/2}(J=2)$\\
3E	&		&	967	&	969.255	&	967.947	&	967.796 	&	$2p_{3/2}^{3}3d_{3/2}(J=1)$\\
3D	&		&	980	&	979.487	&	979.571	&	979.725 	&	$2p_{3/2}^{3}3d_{5/2}(J=1)$\\
3C*	&		&	997	&	997.139	&	997.218	&	997.138 	&	$2p_{1/2}3d_{3/2}(J=1)$\\
b	&	F14	&	1023	&			&	1023.475	&	1023.729 	&	$2p_{3/2}^{2}3d_{5/2}(J=5/2)$\\
	&	F15	&		&			&	1023.475	&	1022.210	&	 $2p_{3/2}^{2}3d_{5/2}(J=3/2)$\\
3B*	&		&	1070	&	1073.451	&			&	1069.006 	&	$2s_{1/2}3p_{1/2}(J=1)$\\
3A*	&		&	1074	&	1079.017	&	1074.844	&	1074.565 	&	$2s_{1/2}3p_{3/2}(J=1)$\\
4G 	&		&	1188	&	1190.006	&			&	1188.479 	&	$2p_{3/2}^{3}4s_{1/2}(J=1)$\\
4F*	&		&	1207	&	1207.353	&			&	1205.816	&	$2p_{1/2}4s_{1/2}(J=1)$\\
4D	&		&	1226	&	1228.023	&	1226.207	&	1226.450	&	$2p_{3/2}^{3}4d_{5/2}(J=1)$\\
4C*	&		&	1243&	1244.359	&			&	1242.799	&	$2p_{1/2}4d_{3/2}^{3}(J=1)$\\
5G	&		&	1322	&	1322.537	&			&	 		&	$2p_{3/2}^{3}5s_{1/2}(J=1)$\\
c	& 	5E	&	1340	&	1338.760	&			&	1338.739	&	$2p_{3/2}^{3}5d_{3/2}(J=1)$\\
	& 	5F*	&		&	1340.036	&			&	 		&	$2p_{1/2}5s_{1/2}(J=1)$\\
	&	5D	&		&	1341.376	&			&	1339.897	&	$2p_{3/2}^{3}5d_{5/2}(J=1)$\\
d	&	4B*	&	1356	&	1356.458	&			&	1354.682	&	$2s_{1/2}4p_{1/2}(J=1)$\\
	&	5C*	&		&	1358.142	&			&			&	$2p_{1/2}5d_{3/2}(J=1)$\\
	&	4A*	&		&	1358.583	&			&	1356.609	&	$2s_{1/2}4p_{3/2}(J=1)$\\
6G	&		&	1390	&	1391.931	&			&	 		&	$2p_{3/2}^{3}6s_{1/2}(J=1)$\\
6D	&		&	1401	&	1402.566	&			&	1401.221 	&	$2p_{3/2}^{3}6d_{5/2}(J=1)$\\
e	&$\dagger$&	1423	&			&			&	 		&						\\
7G	&		&	1432	&	1432.827	&			&	 		&	$2p_{3/2}^{3}7s_{1/2}(J=1)$\\
7D	&		&	1438	&	1439.425	&			&	1439.448 	&	$2p_{3/2}^{3}7d_{5/2}(J=1)$\\
f	&	8E	&	1462	&	1462.704	&			&	1460.988	&	$2p_{3/2}^{3}8d_{3/2}(J=1)$\\
	&	8D	&		&	1463.288	&			&	 		&	$2p_{3/2}^{3}8d_{5/2}(J=1)$\\
g	&	9G	&	1479	&	1476.579	&			&	 		&	$2p_{3/2}^{3}9s_{1/2}(J=1)$\\
	&	8F* 	&		&	1476.658	&			&	 		&	$2p_{1/2}8s_{1/2}(J=1)$\\
	&	9D	&		&	1479.618	&			&	 		&	$2p_{3/2}^{3}9d_{5/2}(J=1)$\\
	&	8C*	&		&	1480.824	&			&	 		&	$2p_{1/2}8d_{3/2}(J=1)$\\
	&	5B*	&		&	1481.395	&			&	 		&	$2s_{1/2}5p_{1/2}(J=1)$\\
	&	5A*	&		&	1482.437	&			&	 		&	$2s_{1/2}5p_{3/2}(J=1)$\\
h	&	10G 	&	1490	&	1489.084	&			&	 		&	$2p_{3/2}^{3}10s_{1/2}(J=1)$\\
	&	10D	&		&	1491.282	&			&	 		&	$2p_{3/2}^{3}10d_{5/2}(J=1)$\\
i	&	9C*	&	1499	&	1497.226	&			&	 		&	$2p_{1/2}9d_{3/2}(J=1)$\\
	&	11G	&		&	1498.262	&			&			&	 $2p_{3/2}^{3}11s_{1/2}(J=1)$\\
	&	11D	&		&	1499.902	&			&	 		&	$2p_{3/2}^{3}11d_{5/2}(J=1)$\\ 
\insertTableNotes
\end{longtable}
\end{ThreePartTable}

Another key diagnostic of CX stems from the fact that CX typically leads to electron capture into a high-$n$ state \citep{1983PhST....3..208J}. The $l$ capture state, while harder to predict, has been shown to vary with the collision energy: at high collision energies, the states are populated statistically, favoring higher angular momentum states \citep{1985PhR...117..265J}, and at the low collision energies produced with the EBIT, lower angular momentum states are favored \citep{1979PhRvA..20.1828R, 2000PhRvL..85.5090B}. This leads to a strong $11s$ or $11d \rightarrow 2p$ transition observed in the Ni$^{19+}$+H$_2$ spectrum and $8d$ or $9d \rightarrow 2p$ transition in the Ni$^{19+}$+He spectrum. The decrease in the $n$\textsubscript{max} state between experiments with H\textsubscript{2} and He likely stems from their differing first ionization potentials ($\sim$15.4 eV and $\sim$24.6 eV, respectively). 

We also observe the presence of lines that result from core-excited states: for example, an electron hole in $2s_{1/2}$ being filled by either a $3p_{3/2}$ or $3p_{1/2}$ electron (to create the 3A and 3B lines, respectively), or a hole in $2p_{1/2}$ being filled by a $3s_{1/2}$ electron (to create the 3F line). This is surprising under the naive assumption that the parent F-like ion would be in the ground state with no core excitation, $1s_{1/2}^{2}2s_{1/2}^{2}2p_{1/2}^{2}2p_{3/2}^{3}$, and that CX would not influence the core configuration. However, previous experiments and subsequent modeling of CX with higher-Z ions (e.g. \cite{Tawara_core, Schuch}) have shown that electron-electron interactions (i.e., configuration mixing) during or after electron capture can lead to core excited states.

\begin{figure}[ht]
 \begin{center}
 \includegraphics[scale=0.7]{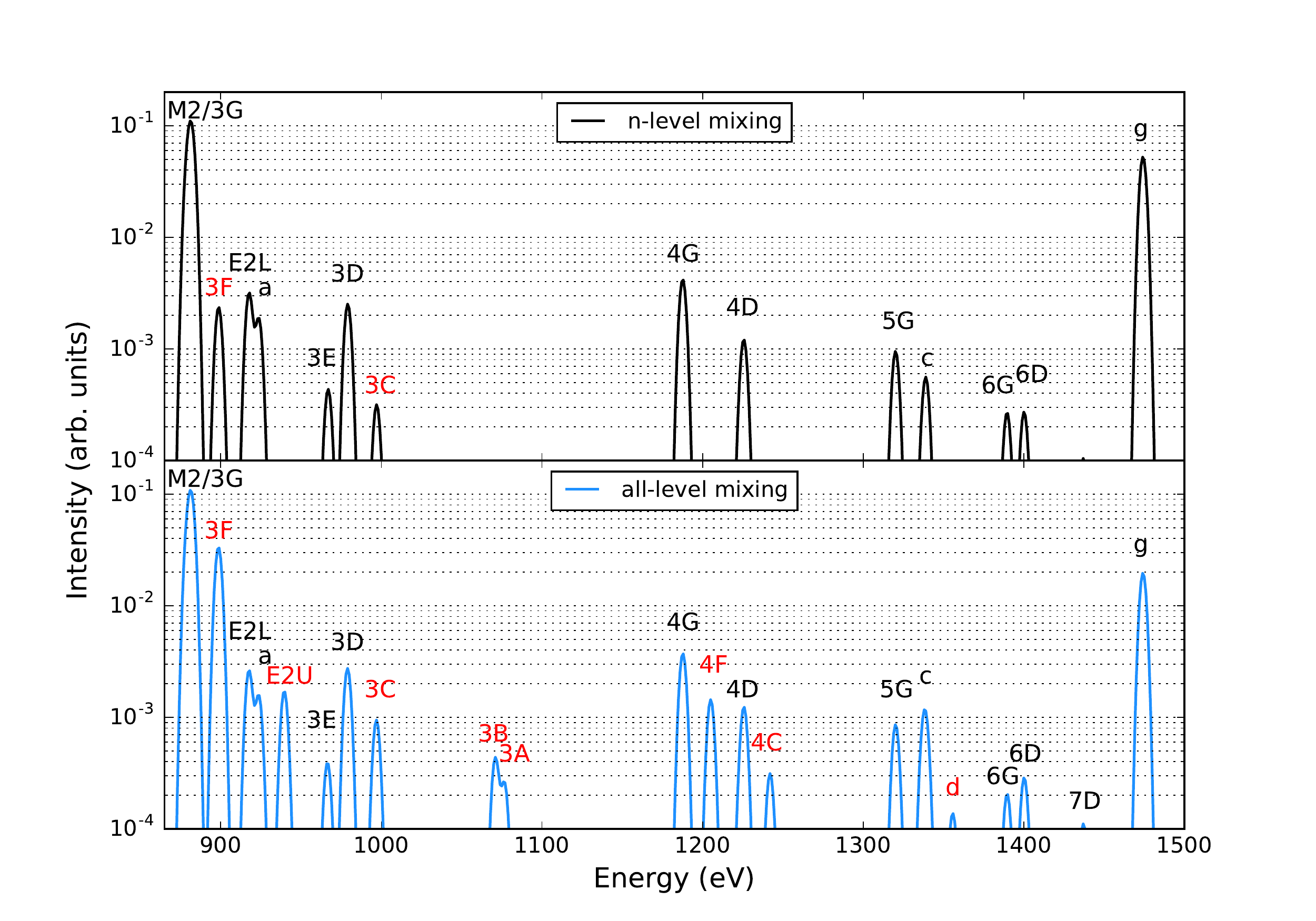}
 \end{center}
 \renewcommand{\baselinestretch}{1}
\small\normalsize
\begin{quote}
\caption{Simulated spectra resulting from cascades following electron capture into $2p_{3/2}^{3}9s_{1/2}(J=1)$ (creating Ne-like Ni), via \textsc{fac} structure calculations, with two different mixing schemes. Top: only levels with like $n$ for the excited electron mix. Bottom: all levels mix. Line labels correspond to entries in Table \ref{table:lineID}, and transitions resulting from core excited states are highlighted in red. For this capture state, allowing all levels to mix is more likely to generate core excited states, and in general, the two mixing schemes generate distinct spectra.}
\label{fig:mixcompare} 
\end{quote}
\end{figure}

To further investigate this effect for our ion of interest, we performed \textsc{fac} structure calculations of Ni$^{18+}$ with one excited electron using two different mixing schemes: mixing only between levels with the excited electron in the same $n$-level, and mixing between all levels. A comparison of cascade spectra resulting from capture into a single excited state is presented in Figure \ref{fig:mixcompare}. We found that lines from upper levels with core excited states are more likely to occur if all levels are allowed to mix, and in general, the two resulting spectra can be dramatically different. This reinforces the results from \cite{Tawara_core} and \cite{Schuch} that mixing is likely the main mechanism for generating core excitation following CX. In addition, this indicates that how mixing is treated in cascades can greatly impact the resulting spectrum, and should be considered carefully. With an adequate understanding of the mixing configurations and subsequent decay schemes, core excited lines following CX may be an additional diagnostic of the quantum state of the captured electron.

\section{Spectral Models}\label{sec:spex_acx}

We used \textsc{spex-cx} version 3.0 and \textsc{acx} version 1.0 to simulate spectra for interactions between Ni$^{19+}$+H (\textsc{spex-cx}, \textsc{acx}) and Ni$^{19+}$+He (\textsc{acx} only) and to compare with our experimental spectra. Although in some cases, the models and data involve differing neutrals, some have similar ionization potentials (IP, 15.4 eV for H\textsubscript{2} and 13.6 eV for H, though a slightly higher IP of 24.6 eV for He), and all have at most two electrons available for capture. We believe that our most significant findings do not depend on the neutral partner, as the main effect we expect is a decrease in $n$\textsubscript{max} with increasing IP \citep{1985PhR...117..265J}. However, multi-electron capture (MEC) must still be considered in the H$_2$ and He cases, and intrinsic differences in the state-selective cross sections between the various neutrals may also be present. 

To simulate CX in \textsc{acx}, we used the \textsc{acxion} model and convolved the resulting spectrum with a Gaussian line profile to match the instrumental response of the ECS. We set the parent ion to be Ni$^{19+}$, and considered CX with either pure H or pure He by adjusting the \texttt{fracHe0} parameter. We used two different $l$-distributions for the captured electrons by varying the \texttt{model} parameter: model ``8," which is the default model used in \textsc{acx} and which assumes a separable $l$-distribution, and model ``15," which uses a Landau-Zener weighting function for the total $L$-distribution. Model 15 was chosen because Landau-Zener methods are known for being most applicable to low-energy CX collisions such as those in EBITs \citep{1985phci.book.....J}. 

For the \textsc{spex-cx} model, we used the lowest allowable collision velocity of 50 km s$^{-1}$, the approximate ionization temperature of 500 eV, and zeroed out the abundance of all ions except Ni. While this does not exactly describe our system---in particular, in the mode in which we operated the electron beam, it does not have a thermal Maxwellian electron distribution---500 eV yielded the closest match to our spectra upon visual inspection after stepping through several values, and the lowest c-statistic after performing a model fit to the data. We left the \texttt{weight} parameter at the default value, which picks the most appropriate $l$-distribution based on the velocity. In our case, this was the Landau-Zener $l$-distribution. We convolved the model spectrum with a Gaussian line profile. \textsc{spex-cx} does not have the ability to consider neutral partners other than H. 

\section{Model Comparison Results}\label{sec:compare}

\begin{figure}
 \begin{center}
 \includegraphics[scale=1.15, angle=90]{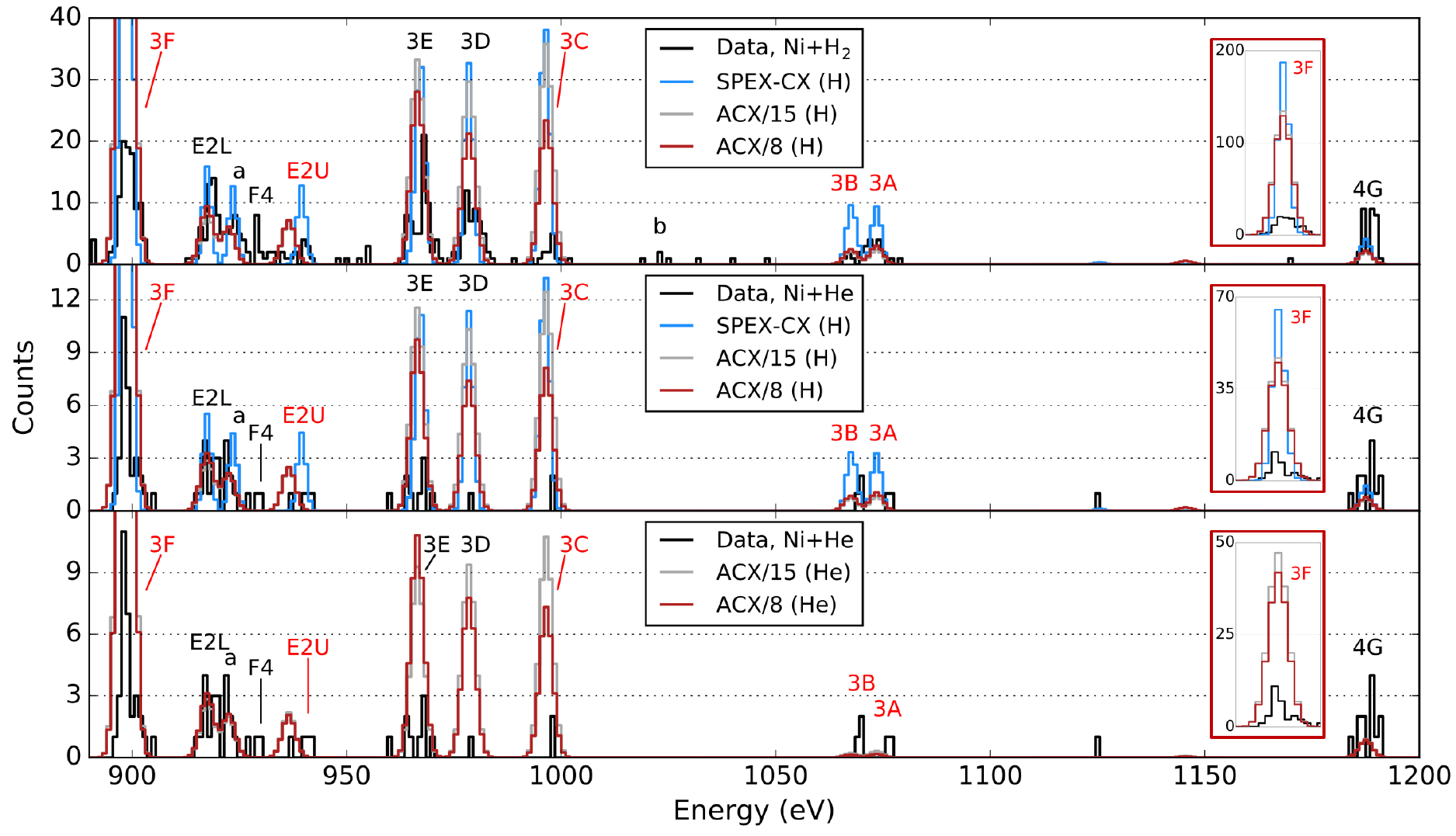}
 \end{center}
 \renewcommand{\baselinestretch}{1}
\small\normalsize
\begin{quote}
\caption{EBIT data compared to the \textsc{spex-cx} and \textsc{acx} models from 890--1200 eV, with the neutral used in each spectrum indicated in the legend. Lines resulting from core excited states are indicated with red labels. Model line strengths are normalized to the total number of counts in the M2/3G lines in the data being compared in each panel (not shown for scale).}
\label{fig:mc1}
\end{quote}
\end{figure}

\begin{figure}
 \begin{center}
 \includegraphics[scale=0.78, angle=90]{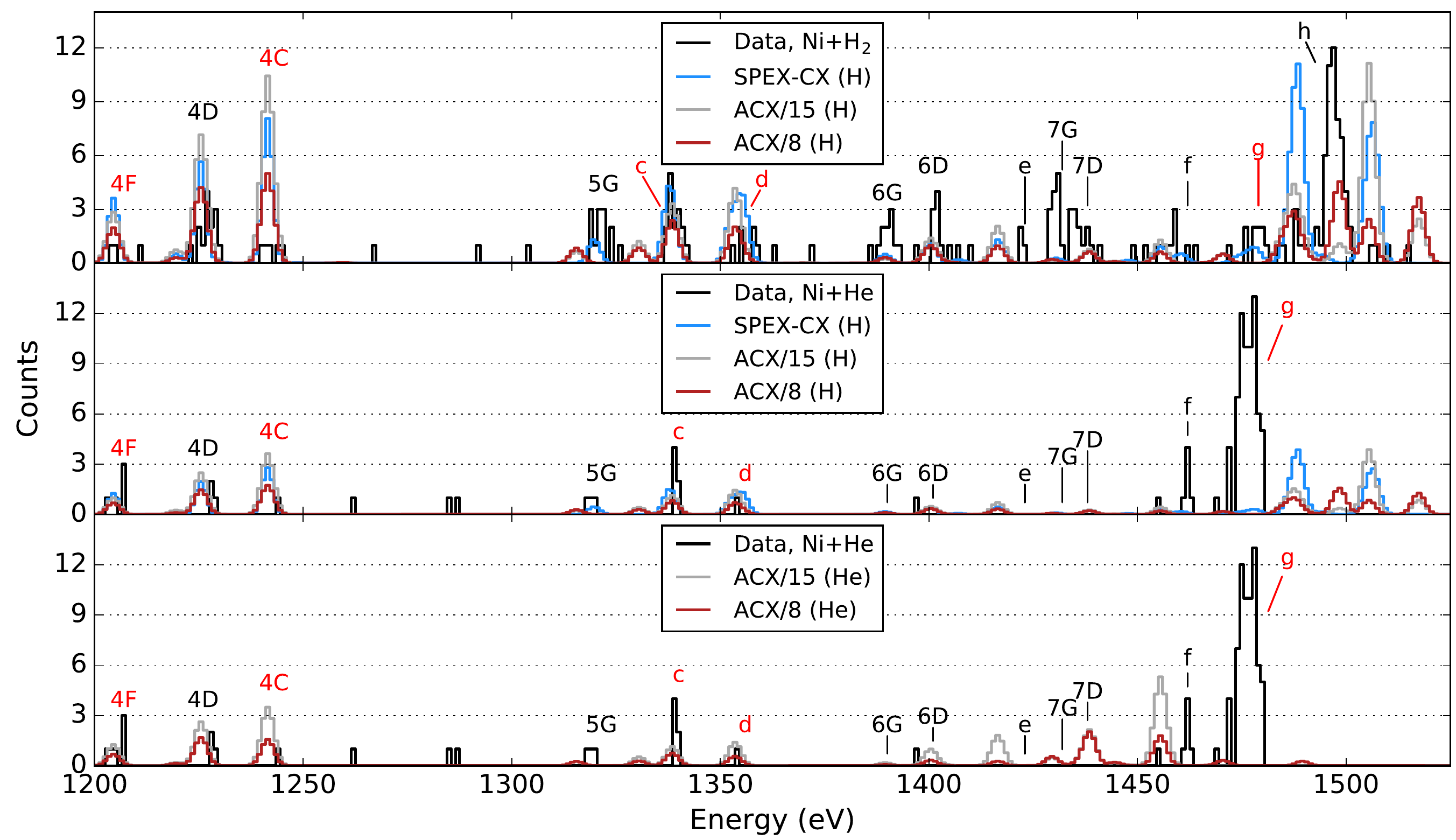}
 \end{center}
 \renewcommand{\baselinestretch}{1}
\small\normalsize
\begin{quote}
\caption{As in Figure \ref{fig:mc1}, but from 1200--1525 eV.}
\label{fig:mc2}
\end{quote}
\end{figure}

Figures \ref{fig:mc1} and \ref{fig:mc2} show the results of the comparisons between models and data, in some cases cross-comparing spectra assuming different neutrals. In general, these comparisons show that although some lines are well approximated by the models, most lines are dramatically over- or under-predicted, and in some cases, line energies for the same atomic transition differ between the models and those identified in our data by over 10 eV.

An important energy regime that the models fail to correctly reproduce is near the strong high-$n \rightarrow n=2$ transition(s) between 1450--1500 eV. The canonical equation to estimate the primary $n$ capture state of the transferred electron, used in both \textsc{spex-cx} and \textsc{acx} codes for CX with L-shell Ni, is:

\begin{equation} \label{eqn:n_max}
n\textsubscript{max} = q\sqrt{\frac{I_\mathrm{H}}{I_\mathrm{n}}}\left(1+\frac{q-1}{\sqrt{2q}}\right)^{-1/2},
\end{equation}
where $I$\textsubscript{H} and $I$\textsubscript{n} are the ionization potentials of hydrogen and the neutral target, respectively, and $q$ is the ion charge \citep{1985PhR...117..265J}.

This equation predicts $n$\textsubscript{max}$=9.6$ for CX with atomic H. The \textsc{acx} models we used for CX with H thus assume 60\% of electron capture into $n=10$ and 40\% into $n=9$. The \textsc{spex-cx} model for CX with H assumes capture into $n=10$.\footnote{The \textsc{spex-cx} model determines the initial $n$ distribution by inputing the calculated $n$\textsubscript{max} value from Equation \ref{eqn:n_max} into Equation A.1 in \citet{2016A&A...588A..52G}. This derives an energy-dependent $n$ distribution, which for our case yields $n$\textsubscript{max}$=10.4$. For ions with odd-numbered charge, this $n$ distribution is then empirically shifted and results in $n=10.7$. However, energy levels for $n>10$ are not available for this ion, so the capture state was set to be $n=10$ (Liyi Gu, private communication).} For CX with He, Equation \ref{eqn:n_max} predicts $n$\textsubscript{max}$=7.1$, and $n$\textsubscript{max}$=9.0$ for CX with H\textsubscript{2}. 

These model estimates clearly differ from our measurements, where there is very little capture into these $n$ levels: the primary capture state in our spectra of Ni$^{19+}+$H\textsubscript{2} is $n=11$, and 8--9 for Ni$^{19+}+$He. This highlights the approximate nature of this equation, which may cause problems when fitting spectra from celestial sources. Furthermore, transitions that involve $n>10$ are not included for any ions in \textsc{acx}, and for Ne-like ions in \textsc{spex-cx}, so it is not currently possible for the models to reproduce our results even with a more accurate $n$\textsubscript{max} distribution (Randall Smith and Liyi Gu, private communication).

In addition, the line energies used in the \textsc{acx} model can differ from those from our \textsc{fac} calculation by a large amount. When the spectral lines in this high-$n$ region are identified according to \textsc{acx} versus \textsc{fac}, the $n$ state predicted for a given line can differ by up to $\Delta n=3$ between the two models (Adam Foster, private communication). This corresponds to a difference in $\sim$10 eV between the two models for the same atomic transition. \textsc{acx} also predicts a line at $\sim$1150 eV corresponding to $2s3d\rightarrow$ ground transitions; this line falls $\sim$17 eV away from where these transitions are predicted to occur from \textsc{fac}.

We also found that lines resulting from core-excited states were mostly over-predicted by both models. These lines are highlighted in red in Figures \ref{fig:mc1} and \ref{fig:mc2}. \textsc{acx} and \textsc{spex-cx} both include mixing in their cascades (Adam Foster and Liyi Gu, private communication), but as we saw in Figure \ref{fig:mixcompare}, the details of exactly how mixing is treated can have a large impact on the resulting spectrum. The discrepancies we observe between our data and models show that more detailed calculations of the strength of configuration mixing and levels involved should be performed, compared to experimental benchmarks, and incorporated into these models to improve their accuracy.


\section{Summary}

We have presented high-resolution spectra of CX between L-shell Ni and neutral H$_2$ and He, and identify spectral diagnostics. We find that configuration mixing is an important effect to include in CX models to generate the core excitation that we observe, as in \cite{Tawara_core, Schuch}. We show that \textsc{spex-cx} and \textsc{acx} models do not accurately reproduce our experimental results, with disconcerting differences across the L-shell energy band. This is likely due to the approximate nature of the scaling equations used to estimate the $n$\textsubscript{max} and $l$ distributions, limitations in the databases for high-$n$ energy levels, inaccurate line energies, and improper treatment of configuration mixing, though intrinsic differences between state-selective cross sections with H, H\textsubscript{2}, and He may also play a role.

While the availability and relative ease of use of models such as \textsc{acx} and \textsc{spex-cx} are beneficial for encouraging the incorporation of CX into spectral analyses, these comparisons show that these models must be used with caution. If they are applied to astrophysical data, incorrect assumptions may be made about the physical and chemical parameters of the observation target, such as the neutral species present, the ion or neutral abundance, or the ion/neutral collision energy, due to potentially misleading predicted line strengths or ratios. The discrepancies we show between experiments and models presented here highlight the need for careful atomic structure calculations for L-shell ions, including both transition energies and mixing coefficients, in order to generate accurate cascades, as well as detailed state-selective cross section calculations for CX, particularly at low collision energies. It is also important to compare these values across models and against laboratory data in order to ensure that spectra from CX and other recombination processes are sufficiently accurate. These calculations and comparisons are especially critical to perform in advance of the forthcoming high-resolution spectra from future satellites such as XRISM and Athena.

\acknowledgments
The authors thank Ed Magee and David Layne at LLNL for their technical support. Work by G. Betancourt-Martinez was partially supported by a NASA Space Technology Research Fellowship. R. Cumbee was supported by an appointment to the NASA Postdoctoral Program at NASA/GSFC, administered by Universities Space Research Association. Work at the Lawrence Livermore National Laboratory was performed under the auspices of the U.S. Department of Energy under Contract DE-AC52-07NA27344 and supported by NASA APRA grants to LLNL and NASA/GSFC.


\end{document}